\documentclass[12pt, draftclsnofoot, onecolumn]{IEEEtran}

\usepackage{microtype}
\usepackage{graphicx}
\usepackage{cite}
\usepackage{booktabs} 
\usepackage{mathtools}
\usepackage{amsmath}
\usepackage{mathrsfs}

\usepackage{amssymb}
\usepackage{bbm}
\usepackage{algorithm}
\usepackage{algorithmic}
\usepackage{amsthm} 
\usepackage{comment}
\usepackage{subcaption}
\usepackage{multirow}
\usepackage{booktabs}

\usepackage{xcolor}

\usepackage{hyperref}
\usepackage{xr-hyper}
\newtheorem{theorem}{Theorem}
\newtheorem{proposition}{Proposition}

\newtheorem{definition}{Definition}
\newtheorem{lemma}{Lemma}

\DeclareMathOperator*{\argmin}{\arg\!\min}

\begin{document}

\title{DRL meets DSA Networks: Convergence Analysis and Its Application to System Design}
\author{Ramin Safavinejad,
        Hao-Hsuan Chang, and
        Lingjia Liu
\thanks{R. Safavinejad and L. Liu are with Wireless@Virginia Tech, Bradley Department of Electrical and Computer Engineering, Virginia Tech. H. Chang is with the Standards and Mobility Innovation Laboratory of Samsung Research America. 
}
}



\maketitle

\begin{abstract}
In dynamic spectrum access (DSA) networks, secondary users (SUs) need to opportunistically access primary users' (PUs) radio spectrum without causing significant interference.
Since the interaction between the SU and the PU systems are limited, deep reinforcement learning (DRL) has been introduced to help SUs to conduct spectrum access. 
Specifically, deep recurrent Q network (DRQN) has been utilized in DSA networks for SUs 
to aggregate the information from the recent experiences to make spectrum access decisions.
DRQN is notorious for its sample efficiency in the sense that it needs a rather large number of training data samples to tune its parameters which is a computationally demanding task.
In our recent work, deep echo state network (DEQN) has been introduced to DSA networks to address the sample efficiency issue of DRQN.
In this paper, we analytically show 
that DEQN comparatively requires less amount of training samples than DRQN to converge to the best policy.
Furthermore, we introduce a method to determine the right hyperparameters for the DEQN providing system design guidance for DEQN-based DSA networks.
Extensive performance evaluation confirms that DEQN-based DSA strategy is the superior choice with regard to computational power while outperforming DRQN-based DSA strategies.

\end{abstract}

\begin{IEEEkeywords}
Deep reinforcement learning (DRL), Covering Numbers, Dynamic Spectrum Access (DSA), Echo State Network (ESN), Recurrent Neural Network, 5G Beyond and 6G
\end{IEEEkeywords}

\section{Introduction}
%
Dynamic spectrum access (DSA) was originally introduced in cognitive radio networks (CRN) to allow secondary users (SUs) to share the spectrum licensed by primary users (PUs) without causing harmful interference to the PUs~\cite{haykin2015cognitive}. 
Through PU-SU coexistence, DSA is regarded as a promising technology to improve the overall spectrum utilization of a CRN significantly~\cite{chakraborty2016design}.
Similar ideas have been extended to allow spectrum sharing and coexistence among different wireless systems.
For instance, through licensed assisted access (LAA)~\cite{Rupasinghe2015reinforce} the 4G LTE-Advanced system is able to coexist with Wi-Fi systems in the unlicensed spectrum, while 4G and 5G systems are able to dynamically share the licensed spectrum~\cite{DSS}.
Meanwhile, coexistence between different wireless systems are introduced to allow different wireless users potentially from different services (e.g., IoT network, 5G, Wi-Fi etc) to share the radio spectrum to improve the spectrum utilization.  
Therefore, shared spectrum access, especially the dynamic spectrum access, that promotes the most effective and efficient spectrum uses has been regarded as one of the key technology enablers for 6G networks~\cite{NGA}. 
%
%
%
%
%
%
%
%
Still, the potential of DSA is yet to be fully realized in practice because of the technological barriers in coping with the complex and highly dynamic radio environments that SU devices have to interact with to gain real-time channel access and interference control.  
This makes the coexistence of primary and secondary systems extremely challenging. 
With the recent surge of artificial intelligence (AI), it becomes possible to employ secondary devices to sense and access the channels smartly in highly dynamic networks. 
Among available AI/machine learning (ML) tools, deep reinforcement learning (DRL) becomes a natural framework to solve for DSA networks~\cite{RubayetAI,chang2018distributive}.
To be specific, with the help of DRL, SU devices will be able to adapt their spectrum access strategies to unknown dynamic radio environments without modeling the underlying environments.
%

On the other hand, developing DRL-based DSA strategies for mobile networks is usually difficult due to the following challenges. 
\begin{itemize}
\item Challenge 1: Partial observability: 
SU devices can be resource constrained IoT devices that are not able to conduct accurate spectrum sensing across all available bandwidth simultaneously.
This will lead to a partially observable radio environment to SU devices with noisy inputs.
%
\item Challenge 2: Limited over-the-air (OTA) training: OTA training data is generally costly in mobile networks since it imposes control overhead~\cite{RubayetAI}.
Meanwhile, the highly dynamic nature of the radio environment also makes the collected OTA training data outdated in a relatively short time.
\end{itemize}

%
%
Combining recurrent neural network (RNN) and deep Q-network (DQN), the deep recurrent Q-network (DRQN), has been introduced as a DRL technique to aggregate information from past observations in a partially observable Markov decision process (POMDP) environment~\cite{hausknecht2015DRQN}.
Action-specific Deep Recurrent Q-Network (ADRQN) \cite{igl2018deep} further enhances DRQN by using both observation and action as inputs of RNNs.
Therefore, both techniques have been introduced to enable DRL-based DSA to address Challenge 1~\cite{YueDSA}.
However, the training of the underlying RNN of both methods suffers from poor convergence requiring a lot of training samples~\cite{2013RNNdifficulty}, which is infeasible in a highly dynamic environment with limited OTA training data ----- Challenge 2.
Meanwhile, generative models of the environment have been introduced at the DRL agents to allow them to learn and to infer the belief states using the learned models~\cite{igl2018deep,huang2019svqn}.
However, these methods are not suitable for DSA networks since the learned generative model will become outdated quickly in such a highly dynamic environment.
Therefore, it is of critical importance to design a fast-converging DRL strategy that can work efficiently in a partially observable and highly dynamic environment to jointly address both challenges of the DRL-based DSA strategies for mobile networks.

In light of the challenges, a new DRL strategy called deep echo state Q network (DEQN) has been introduced in~\cite{Hao-Hsuan-DEQN}.
Echo state network (ESN)~\cite{jaeger2001echo} has been adopted in DEQN as the kernel of the Q-network to reduce the training time and the required training data.
Note that ESN is a special type of RNN that only trains the output weights while leaving input weights and recurrent weights untrained.
%
%
ESNs have been shown to achieve comparable performance with RNNs, especially in some applications requiring fast learning~\cite{tanaka2019RC}.
%
Furthermore, the hidden states of ESNs are unchanged during the training process due to the fact that both input weights and recurrent weights are fixed.
In this way, DEQNs can pre-store hidden states, while DRQNs have to recalculate hidden states in every training iteration.
Accordingly, DEQNs can substantially decrease the training time and the amount of training data.

In this paper, we will focus on DEQN to show that it can be used to address both of the challenges for DRL-based DSA strategies and introduce a method to determine the right hyperparameters for the DEQN providing system design guidance for DEQN-based DSA schemes.
This is done through the notion of covering numbers, which capture the complexity of a given function class such as that of RNNs or ESNs.
Covering numbers portray this complexity by determining the least number of hyperballs of radius $\epsilon$ we would need to cover the output space of a function class given a fixed input set.
If a function class needs more hyperballs for it to be covered, then it is a richer and, consequently, more complex function class.
It is shown in the literature that the covering number of a Q-network can upper bound a notion called the one step approximation error~\cite{fan2020theoretical}, which shows how fast DRL would converge with that specific Q-network as its kernel.
For the theoretical convergence analysis of this paper, we first upper bound the covering number of RNNs and ESNs, which is outlined in Section \ref{sec:theoretical}.
We then use this upper bound on the covering number to characterize the convergence rate.
The work in~\cite{RNNchen} provides an upper bound on the covering number of a few variations of RNNs. 
However, that work does not consider the effect of the number of training sequences that are used for training.
Therefore, we cannot make use of the upper bound that is obtained in~\cite{RNNchen} for our work.

%



In summary, the main contributions of our work are as follows: 
%
%
\begin{enumerate}
    \item An upper bound has been obtained on the covering number of single layer RNNs and ESNs.
    \item Theoretical convergence analysis of both DEQN and DRQN have been conducted by means of covering numbers to demonstrate the faster convergence rate of DEQN compared to DRQN.
    \item A bias-variance trade-off has been characterized for DEQN where the variance term is characterized by its covering number and the bias term is determined empirically through a quadratic optimization problem.
    \item A method has been introduced for determining the spectral norm of the recurrent weights of the ESN kernel of DEQN to minimize the bias plus variance error. It will provide theoretical foundation for choosing hyperparameter for the underlying DEQN-based DSA strategies. 
    \item Extensive performance evaluation has been conducted for both DEQN and DRQN on a highly dynamic DSA environment to demonstrate the benefits of DEQN-based DSA strategy over its DRQN counterpart.
\end{enumerate}

%
%
%
%

%

%


In what follows, we will go over the DRL problem formulation, introduction of DEQN and its training method in Section \ref{sec:DEQN_introduction}, provide theoretical analysis of DEQNs and DRQNs in Section \ref{sec:theoretical}, formulate the 5G DSA problem in Section \ref{sec:DSA_formulation}, showcase numerical results and performance evaluation in Section \ref{sec:performance} and finally, conclude the paper in Section \ref{sec:conclusion}.


\section{Deep Echo State Q Network}
\label{sec:DEQN_introduction}
In this section we first formally state the reinforcement learning (RL) problem for our theoretical analysis and then introduce the DEQN's update equations and training algorithm. 
\subsection{RL Problem Setup}
For the theoretical analysis, we assume the underlying environment has the Markov property and the problem can be formulated as a Markov Decision Process (MDP).
we define the problem as an MDP rather than a POMDP, due to how general a POMDP can be.
In the literature, theoretical analysis of POMDPs is mostly done with knowledge of the model and without learning based approaches (see e.g.~\cite{POMDPservey}).
However, we will also apply the introduced learning methods to partially observable environments in our performance evaluation.
An MDP model is characterized by a tuple ($\mathcal{S}$, $\mathcal{A}$, $\mathcal{T}$, $R$, $\gamma$), 
where $\mathcal{S}$, $\mathcal{A}$, $\mathcal{T}$, $R$ and $\gamma \in (0, 1)$ are the state space, action space, state transition probability, reward function and discount factor for future rewards respectively.
To put it in more detail, the agent observes state $s_t \in \mathcal{S}$ at time $t$ and takes the action $a_t \in \mathcal{A}$. 
After taking the action $a_{t}$, the agent receives a reward $r_t= R(s_{t}, a_{t})$, 
and then the system shifts to the next state $s_{t+1}$ based on the state transition probability $\mathcal{T}(s_{t+1} |s_{t}, a_{t})$.
The the agent's goal is to find a policy $\pi$ for making a sequence of action decisions that maximize the expected discounted reward, $\mathbb{E}_{\pi} \left[ \sum_{t=1}^{\infty} \gamma^{t-1} r_{t} \right]$.
Also, denote $\Omega_{\le T}$ as the space spanned by $\{s_1,s_2,\dots,s_T\}$.

For the practical implementation setup, we know that the environment is, in general, a POMDP.
Therefore, the agent only has a partial observation of the environment, rather than the full state of it.
However, we will still denote the observation of the agent as $s_t$.
We avoid introducing another notation for the observation as to not confuse the reader when comparing the theoretical analysis with the practical implementation.
The rest of the setup is the same as the theoretical RL setup.

%

%

\subsection{Update Equations of DRQN/DEQN}
The action-value function in DRQN/DEQN is approximated by the Q-network $Q_{\theta}$ with parameter $\theta = \left( W_{\text{in}}, W_{\text{rec}}, W_{\text{out}} \right)$,
where $W_{\text{in}} \in \mathbb{R}^{d_h \times d_x}$ is the input weights,
$W_{\text{rec}} \in \mathbb{R}^{d_h \times d_h}$ is the recurrent weights,
and $W_{\text{out}} \in \mathbb{R}^{d_y \times d_h}$ is the output weights.
A sequence of observations $s_1, \ldots, s_t \in \mathbb{R}^{d_x}$ is input to the Q-network to generate a sequence of outputs $y_1, \ldots, y_t \in \mathbb{R}^{d_y}$ as follows:
\begin{equation}
\begin{aligned}
h_t &= \sigma_h \left( W_{\text{in}} s_t + W_{\text{rec}} h_{t-1} \right) \in \mathbb{R}^{d_h}, \\
y_t &= \sigma_y\left(W_{\text{out}} h_t\right)  \in \mathbb{R}^{d_y},
\label{eqn:update_eqn}
\end{aligned}
\end{equation}
where $\sigma_h$ and $\sigma_y$ are the recurrent  and output units activation functions,
and $h_t$ is the hidden state.
$\sigma_h$ is often chosen as ReLU or tanh activation functions,
and $\sigma_y$ is chosen as the identity function in the case of ESNs.
The hidden state $h_t$ is used to represent a summary of the past sequence of observations up to $t$ and we set $h_0 = \boldsymbol{0}$.
Note that the $a^{\text{th}}$ element of $y_t$ is equal to the estimated action-value function of selecting action $a$, i.e., $y_t^a = Q_{\theta} \left(s_{\leq t}, a \right)$, where $s_{\leq t} = \left( s_1, \ldots, s_t \right)$.
We define the action-value function as the sum of discounted rewards from the current time step until the future, given that the sequence of states until now is $s_{\leq t}$ and the action taken is $a_k$, in other words
\begin{align}
    Q_{\theta} \left(s_{\leq t}, a \right) = \mathbb{E}\Bigg[\sum_{\tau=0}^{\infty} {\gamma^{\tau}r_{t+\tau}} \bigg| S_0=s_0,\dots,S_t=s_t,A_t=a_t \Bigg]
\end{align}
where $r_t$ is the reward received at time step $t$. 
We also define the operator $\mathcal{B}$ as the Bellman operator given as
\begin{align}
    \mathcal{B} Q_{\theta}(s_{\leq t}, a_t) = r_t + \gamma \max_{a} Q_{\theta}(s_{\leq t+1}, a)
\end{align}
If for a policy $\pi_L^*$ we have $\mathcal{B} Q_{\theta}^{\pi_L^*}(s_{\leq t}, a_t) = Q_{\theta}^{\pi_L^*}(s_{\leq t}, a_t) $, we call $\pi_L^*$ an optimal policy and define $Q^* = Q^{\pi_L^*}$.

%
We therefore define the loss function for training $\theta$ as 
\begin{equation}
    \label{eqn: loss}
    \left( r_t + \gamma \max_{a} Q_{\theta^{-}}(s_{\leq t+1}, a) - Q_{\theta}(s_{\leq t}, a_t) \right)^2,
\end{equation}
where $\theta^{-}$ is the parameter of a target Q-network initially identical to the main Q-network.
To stabilize the training, $\theta^{-}$ is periodically synchronized with $\theta$ instead of being updated in each training iteration. 
\subsection{Training Algorithm for DEQN}
BPTT involves unfolding the network in time into a computational graph that has a repetitive structure, which suffers from the slow convergence rate due to vanishing and exploding
gradients \cite{2013RNNdifficulty}.
Furthermore, DRQN requires a large amount of training data to ensure the learning agent converges to an appropriate policy.
Unfortunately, in the DSA problem for 5G, the DRL agent can only collect limited training data and the 5G wireless environment is highly time varying. 
To quickly adapt to such an environment using limited training data, we adopt the online training algorithm for DEQN originally introduced in \cite{Hao-Hsuan-DEQN}.

\begin{algorithm}[ht]
\caption{The online training algorithm for DEQN.}
\begin{algorithmic}



\STATE {
Initialize a Q-network $\theta$ and a target Q-network $\theta^{-}$.
}

\FOR {$l = 1, ..., L$}

\STATE {
Set $\theta^{-} = \theta$ and empty the replay buffer $D$.
}

\STATE {
Observe $o_1$ from the environment.
}

\FOR {$t = 1, ..., T$}

\STATE {
Input $s_t$ and $h_{t-1}$ to $Q_{\theta}$ to calculate $h_{t}$ and $y_t$.
}


\STATE {
Execute $a_t$ based on $y_t$ and then receive $r_t$.
}

\STATE {
Observe $s_{t+1}$ from the environment. 
}

\STATE {
Input $s_{t+1}$ and $h_{t}^{-}$ to $Q_{\theta^{-}}$ to calculate $h_{t+1}^{-}$.
}

\STATE {
Store ($h_{t}$, $a_t$, $r_t$, $h_{t+1}^{-}$) in replay buffer $D$.
}

\ENDFOR

\STATE {
Sample ($h_{j}$, $a_{j}$, $r_{j}$, $h_{j+1}^{-}$) from replay buffer $D$.
}


\STATE {
Update $\theta$ by performing gradient descents on \begin{center} $\left( r_j + \gamma \max\limits_{a} Q_{\theta^{-}}(h_{j+1}^{-}, a) - Q_{\theta}(h_{j}, a_j) \right)^2$. \end{center}
}

\ENDFOR

\end{algorithmic}
\label{alg:DEQN_training}
\end{algorithm}

The designed DEQN training algorithm is stated in Algorithm~\ref{alg:DEQN_training}.
First, the input weights and the recurrent weights of ESNs are initialized randomly, and then they remain untrained.
Only the output weights of ESNs are trained so the training is extremely fast by avoiding BPTT.
Second, given $(s_1, \ldots, s_t)$, we can observe that the hidden states $(h_1, \ldots, h_t)$ are unchanged during the training process from (\ref{eqn:update_eqn}) because the input weights and recurrent weights are fixed.
In contrast to DRQNs that waste some training samples and computational resources to recalculate the hidden states in every training iteration, DEQNs can pre-store the hidden states in the replay buffer and use them for training.
Therefore, the DEQN training is much more sample and computationally efficient than the DRQN training.
Third, the DEQN training can randomly sample from the replay buffer to create a training batch because the hidden states are unchanged and pre-stored.
On the other hand, the DRQN training has to sample continuous sequences to create a training batch.
Breaking the temporal correlations of sampled data during training is crucial for reducing generalization error, as the stochastic optimization algorithms usually assume i.i.d. data.
Lastly, to deal with the highly dynamic and time varying environment, the outdated training data in the replay buffer will be removed and the policy will be updated continually by the effective data collected from the latest environment.

\section{Theoretical Analysis for DRQN and DEQN}
\label{sec:theoretical}
In this section we will theoretically study the convergence properties of DEQN and DRQN, compare them with each other and provide a method for designing an ESN structure that optimizes the performance of a DEQN.

For the theoretical analysis, we consider the problem of batch reinforcement learning, which is normally considered in the literature for theoretical analysis of RL problems (e.g. \cite{fan2020theoretical,RademacherRL}).
In batch RL, the agent has had a multitude of transitions in the environment and then the training of the Q-network is done after all the transitions have finished.
The agent would sample a batch of size $n$ from those transitions and will do one iteration of training on the neural network using those $n$ transitions.
It is also assumed that sampling from this accumulation of transitions is the same as sampling a transition from the environment with a fixed distribution $\sigma$.
Note that, in our case, the agent collects streams of experiences over the course of multiple episodes and stores them in the experience replay buffer. 
We assume that each episode consists of a sequence of size $T$ of actions taken in the environment.
We denote $\{s_{i,1},s_{i,2},\dots,s_{i,T}\}_{i=1}^{n}$ as a batch of $n$ i.i.d sample sequences that will be used for one training iteration. 
The training continues for another i.i.d batch of sequences for a total of $L$ iterations.
Algorithm \ref{alg:NFQI} outlines the full training procedure that will be used in this section and is regarded as the Neural Fitted Q-Iteration Algorithm. 
We refer the reader to \cite{fan2020theoretical} for a detailed explanation on why this algorithm effectively characterizes Deep Q-Learning.

\begin{algorithm}[ht]
\caption{Neural Fitted Q-Iteration Algorithm.}
\begin{algorithmic}




\STATE {
Initialize $T$ Q-networks with parameters $\{\theta_0^{(t)} \in \Theta\}_{t=1}^{T}$.
}

\FOR {$t=1,\dots,T$}

\FOR {$l = 1, ..., L$}

\STATE {
Sample $n$ sequences of size $T$ of transitions in the environment.
}

\STATE {
Store  $\left\{( s_{i,t}, a_{i,t}, r_{i,t}, s_{i,t+1} \right)\}_{i=1}^{n}$ in replay buffer $D$.
}

\STATE {
Update the action-value function: \\
$\theta_{l}^{(t)} \leftarrow \argmin\limits_{\theta_l^{(t)} \in \Theta} \frac{1}{n} \sum_{i=1}^{n} \left( z_{i,t} - Q_{\theta_l^{(t)}} \left(s_{i,\leq t}, a_{i,t} \right) \right)^2$, \\
where $z_{i,t} = r_{i,t} + \gamma \max_{a} Q_{\theta_{l-1}^{(t)}} \left(s_{i,\leq t+1}, a \right)$.
}

\ENDFOR

\STATE {
Define $\pi_{L}^{(t)}$ as the greedy policy with respect to $Q_{\theta_L^{(t)}}$.
}

\ENDFOR

\end{algorithmic}
\label{alg:NFQI}
\end{algorithm}

Before proceeding with the analysis, we first introduce the notion of covering number, which is widely known in the statistical learning theory community.

\begin{definition}
    Let $\mathcal{F}$ be a function class defined from $\mathbb{R}^{d_x}$ to a metric space $B$ with metric $\normalfont\text{dis}(\cdot)$. 
    For a given set of sample observations $\{ \normalfont\textbf{x}_1,\dots, \normalfont\textbf{x}_n \}$,
    the $\epsilon$-covering number of this function class is denoted by $\mathcal{N} \left(\mathcal{F},\epsilon,\normalfont\text{dis}(\cdot) \right)$ and is defined as the least cardinality $m$ of a set of vectors $\normalfont{V}_m=\{\normalfont\textbf{v}_i:\normalfont\textbf{v}_i \in B^n,i\in[m]\}$, such that for any $f\in \mathcal{F}$ there exists $\normalfont\textbf{v} \in \normalfont{V}_m$ such that
    \begin{align}
        \left( \sum_{i=1}^{n} \normalfont\text{dis} \left( f( \normalfont\textbf{x}_i ), \normalfont\textbf{v}_{(i)}\right)^2 \right)^{1/2}
        \le \epsilon \nonumber
    \end{align}
    where $\normalfont\textbf{v}_{(i)}$ is the $i$-th element of $\normalfont\textbf{v}$
\end{definition}

The above definition is not the most general form of Covering Number definition, but is sufficient for our analysis.
The reader is referred to \cite{zhang2002} and references therein for more general forms of definitions.

In order to analyze the convergence properties of a DQN, we make use of the following two theorems from \cite{fan2020theoretical}:

\begin{theorem} (Theorem 6.1 in~\cite{fan2020theoretical})
Let $\sigma$ be the sampling distribution over $\Omega_{\leq T} \times \mathcal{A}$ in Algorithm \ref{alg:NFQI}, $\mu$ be a fixed probability distribution over $\Omega_{\leq T} \times \mathcal{A}$, and $R_{\text{max}}$ be the maximum reward value.
Then we have
\begin{equation}
\label{eqn:Q_theory_1}
\mathbb{E}_{\mu}\left[ \big|Q^{*} - Q^{\pi_L} \big|  \right] \leq \frac{2 \phi_{\mu,\sigma} \gamma}{(1 - \gamma)^2} \cdot \eta_{\text{max}, T}+ \frac{4 \gamma^{L+1}}{(1 - \gamma)^2} \cdot R_{\text{max}},
\end{equation}
where $\phi_{\mu,\sigma}$ is the concentration coefficient of $\mu$ and $\sigma$, $\pi_{L}$ is the output policy of Algorithm~\ref{alg:NFQI}, and $\eta_{\text{max}, T}$ is the maximum one-step approximation error given by $\eta_{\text{max}, T}=\max_{t\in T} \|\mathcal{B}Q_{t-1}-Q\|$, where $\mathcal{B}(\cdot)$ denotes the Bellman optimality operator.
\end{theorem}
\begin{theorem} 
\label{thm:DRL_bias_variance}
(Theorem 6.2 in~\cite{fan2020theoretical})
Let $\epsilon > 0$ and $V_{\text{max}} = R_{\text{max}}/(1-\gamma)$.
From Theorem 6.2 in~\cite{fan2020theoretical}, the upper-bound of $\eta_{\text{max}, t}$ can be written as
\begin{equation}
\begin{aligned}
\label{eqn:Q_theory_2}
\eta_{\text{max}, T}^2 \leq & 4 \sup_{\theta' \in \Theta} \inf_{\theta \in \Theta} \mathbb{E}_{\sigma}\left[ \left( \mathcal{B} Q_{\theta'} - Q_{\theta} \right)^2 \right]
+ C_1 \cdot V_{\text{max}}^2 / T \cdot \text{log}N_{\epsilon, T} + C_2 \cdot V_{\text{max}} \cdot \epsilon,
\end{aligned}
\end{equation}
where $\mathcal{B}$ is the Bellman optimality operator, $N_{\epsilon, t}$ is the $\epsilon$-covering number of RNN/ESN with respect to the Euclidean norm, and $C_1$ and $C_2$ are constants given by
\begin{equation}
    \begin{aligned}
        &C_1 = \left(8\sqrt{2T} + 256/V_{max}\right)
        ~,~C_2 = \left( 4\sqrt{2n} + 52 \right) 
    \end{aligned} \nonumber
\end{equation}
\end{theorem}
The maximum one step approximation error characterizes the distance between the Q-value at a time instant and the Bellman optimality operator operating on the Q-value of the previous time instant.
The smaller this distance, the closer the Q-network is to approximating the actual Q-values of the environment.
We denote the first term in the RHS of~\ref{eqn:Q_theory_2} as the bias term, and the second and third terms as the variance term.
In what follows next, we derive an upper bound on the covering number which characterizes the variance term, and provide a method for evaluating the bias term.

\subsection{Upper bounding the covering number of RNN/ESN}
We define the class of single layer RNN functions with sequence length $t$ as all the functions that do the operation stated in (\ref{eqn:update_eqn}), their output is $y_t$ and have the matrices $W_{in}$, $W_{rec}$ and $W_{out}$ as trainable parameters but with the constraints
\begin{align}
    \label{eqn: normConstraints}
    \|W_{in}\|_F \le B_{in} , \|W_{rec}\|_F \le B_{rec} , \|W_{out}\|_F \le B_{out}
\end{align}

We denote this class as $\mathcal{H}_{RNN,t}$. Similarly, we denote the class of single reservoir ESN functions with sequence length $t$ as $\mathcal{H}_{ESN,t}$. It is defined exactly the same way as $\mathcal{H}_{RNN,t}$ except for the fact that $W_{in}$ and $W_{rec}$ are fixed and each function in this class is distinguished by $W_{out}$ only.
Note that the constraints in (\ref{eqn: normConstraints}) hold in this class as well.
We also pose a constraint on the input states and let $\|s_i\|\le B_X; i\in [T]$.
\begin{proposition}
\label{prop: RNNcover}
The $\epsilon$-covering number of a single layer RNN is upper bounded by
\begin{align}
    \label{eqn: RNNcover}
    &\ln\left( \mathcal{N}(\mathcal{H}_{RNN,t},\epsilon,\|\cdot\|) \right)
    \nonumber\\
    &\le
    \left( \dfrac{9\rho_y^2 \rho_h^2 B_{out}^2 B_{in}^2 B_X^2 \left( \rho_h^t B_{rec}^t -1 \right)^2}{\left( \rho_h B_{rec} -1\right)^2 \epsilon^2} \right)
    \ln(2d_y d_h)
    \nonumber \\
    &+\dfrac{9t^2 \rho_y^2 \rho_h^2 B_{out}^2 B_{in}^2B_X^2 B_{rec}^2 \rho_h^{2} }{\epsilon^2}
    \left( \dfrac{\rho_h^{2t}B_{rec}^{2t} -1}{\rho_h^{2}B_{rec}^{2} -1} \right)
    \ln(2d_h d_x)
    \nonumber \\
    &+ \dfrac{9(t-1)^2 \rho_y^2 \rho_h^2 B_W^2 B_X^2 B_{out}^2 }{ \epsilon^2 \left( \rho_h B_{rec} -1 \right)^2} 
    \bigg( (t-1)(\rho_h B_{rec})^{2t} -2 \rho_h^{t+1}
    \nonumber \\
     &\cdot B_{rec}^{t+1} \dfrac{(\rho_h B_{rec})^{t-1} -1}{\rho_h B_{rec} -1}
    +\rho_h^{2}B_{rec}^{2} \dfrac{(\rho_h B_{rec})^{2t-2} -1}{\rho_h B_{rec} -1}
    \bigg)
    \ln(2d_h^2)
\end{align}
\end{proposition}

\textit{Proof}: See Appendix \ref{apx: RNNproof} for a detailed proof.

\begin{proposition}
\label{prop: ESNcover}
The $\epsilon$-covering number of a single reservoir ESN is upper bounded by
\begin{align}
    \label{eqn: ESNcover}
    &\ln\left( \mathcal{N}(\mathcal{H}_{ESN,t},\epsilon,\|\cdot\|) \right)\nonumber\\
    &\le
    \dfrac{\rho_y^2 \rho_h^2 B_{in}^2 B_X^2 B_{out}^2 d_y}{\epsilon^2} \left( \dfrac{\rho_h^t B_{rec}^t -1}{\rho_h B_{rec} -1} \right)^2\ln (2d_h d_y)
\end{align}
\end{proposition}

\textit{Proof}: Refer to Appendix \ref{apx: ESNproof} for a detailed proof.

The first term in (\ref{eqn: RNNcover}), which gets multiplied by $\ln (2d_y d_h)$, corresponds to the output weight $W_{out}$.
The other two terms that get multiplied by $\ln (2d_h d_x)$ and $\ln (2d_h^2)$ correspond to $W_{in}$ and $W_{rec}$ respectively.
As can be observed, the upper bound on the ESN's covering number in (\ref{eqn: ESNcover}) is the same as the first term in the RNN's covering number bound in (\ref{eqn: RNNcover}) up to a constant factor.
The other two terms do not appear in the ESN covering number upper bound.
This is because the output weights $W_{out}$ are the only parameters that are trained in the ESN and $W_{in}$ and $W_{rec}$ remain untrained.
This has reduced the richness of the ESN function class compared to RNN, and thus caused a reduction in its covering number.
This reiterates that ESNs are a less complex function class compared to RNNs.

The upper bounds on the covering number allow us to upper bound the variance term in (\ref{eqn:Q_theory_2}) (i.e. the last two terms).
This means that if two Q-network kernels have comparable performance in terms of their best performance, then the one that has a higher covering number will need more training data for the RL algorithm to converge.
In other words, the convergence rate of the one with higher covering number will be lower.
This is a critical property that a Q-network kernel would need in order to adapt to fast-changing environments. 
\subsection{Empirical derivation of the bias term}
\label{subsection:empirical-bias}
The bias term
in~(\ref{eqn:Q_theory_2}) represents how good the Q-network can approximate any function in the function space created by the Bellman operator operating on all the Q-functions formable by the Q-network.
The bias term is highly dependent on the RL environment and it is usually very difficult to find an upper bound on.
For example, the work in~\cite{Ganon2020} provides an upper bound on the average error we would get with a particular way of initializing an ESN.
However, that work requires prior knowledge on the function we are trying to approximate, i.e. the Q-function in our case. 
Specifically, since the Fourier transform of the Q-function of an the RL environment is unknown to us, we cannot make use of the upper bounds that are mentioned there. 

Therefore, we propose to find the bias term empirically during the first episode of the RL training.
We will show in this section that the infimum of the bias error of an ESN can be formulated as a quadratic optimization problem.
It is worth mentioning that such a straightforward approach is not available for RNNs.
The reason for this is that in an RNN there are multiple non-linearities that prevent the optimization problem from becoming even convex.
For the case of ESNs, however, only the output weights are trainable.
This means that the output $y_t$ in (\ref{eqn:update_eqn}) is just a linear combination of the non-linearly derived $h_t$, which for a given set of sample states, is constant.

According to Algorithm \ref{alg:NFQI}, the training update step for the DEQN can be expressed as 

\begin{align}
    \label{eqn:training update}
    \theta_{\text{new}} \leftarrow \argmin\limits_{\theta \in \Theta} 
    \sum_{i=1}^{n} \left( z_{i} - Q_{\theta} \left(\textbf{h}_i, a_{i} \right) \right)^2
\end{align}
where $\theta$ is the parameters of our ESN output weights, 
$n$ is the number of experience replay samples used for one training step, 
$\textbf{h}_i$ is the output of the reservoir for the $i^{\text{th}}$ sample, 
$a_i$ is the action taken for the $i^{\text{th}}$ experience replay sample 
and $z_i$ is defined as
\begin{align}
    z_i = r_i +  \gamma \max_a Q_{\theta_{\text{old}}} (\textbf{h}_i^{'} , a)
\end{align}
where $r_i$ is the reward received in the $i^{\text{th}}$ sample, $\theta_{old}$ is the ESN parameters of the previous training iteration, which in this case is just the initial parameters, and $\textbf{h}_i^{'}$ is the reservoir output of the next state.

Let's consider the optimization problem in (\ref{eqn:training update}).
Substituting the Q function with the output of the ESN shown in (\ref{eqn:update_eqn}), we get
\begin{align}
    \label{eqn: optimization}
    \argmin\limits_{W_{\text{out}}} 
    \sum_{i=1}^{n} \left( W_{\text{out},a_i} \textbf{h}_i - z_{i} \right)^2
\end{align}
where $W_{\text{out},a_i}$ is the row in $W_{\text{out}}$ that leads to the Q value of action $a_i$ when applied to the reservoir output $\textbf{h}_i$.
If we assume $a_i \in \{ 1,2,\dots , |\mathcal{A}| \}$, the optimization in (\ref{eqn: optimization}) can be rewritten as
\begin{align}
    \label{eqn: decomposed optimization}
    \argmin\limits_{W_{\text{out}}} 
    \sum_{k=1}^{|\mathcal{A}|}\sum_{j=1}^{n_k} \left( W_{\text{out},k} \textbf{h}_{k,j} - z_{k,j} \right)^2
\end{align}
where $n_k$ is the number of samples in the experience replay that had led to action $a_k$, $\textbf{h}_{k,j}$ is the $j^{\text{th}}$ reservoir output among the outputs that had led to action $a_k$ and $z_{k,j}$ is defined similarly.
Since the optimization parameters $W_{\text{out},k}$ are mutually exclusive in $k$, we have
\begin{equation}
\label{eqn: loss decomposition}
\begin{aligned}
    \min_{W_{\text{out}}} 
    \sum_{k=1}^{|\mathcal{A}|}\sum_{j=1}^{n_k} \left( W_{\text{out},k} \textbf{h}_{k,j} - z_{k,j} \right)^2
    \\
    =
    \sum_{k=1}^{|\mathcal{A}|}\min_{W_{\text{out,k}}} \sum_{j=1}^{n_k} \left( W_{\text{out},k} \textbf{h}_{k,j} - z_{k,j} \right)^2
\end{aligned}
\end{equation}
Therefore, it is equivalent to minimize each of the inner objective functions separately.
In other words
\begin{align}
    {W_{\text{out,k}}^*} = \argmin_{W_{\text{out,k}}} \sum_{j=1}^{n_k} \left( W_{\text{out},k} \textbf{h}_{k,j} - z_{k,j} \right)^2
\end{align}
Let $L_k = \sum_{j=1}^{n_k} \left( W_{\text{out},k} \textbf{h}_{k,j} - z_{k,j} \right)^2$ and denote $\textbf{w}_k = W_{\text{out},k}^{T}$. We have
\begin{align}
    \label{eqn: quadratic}
    \min_{\textbf{w}_k} L_k = \min_{\textbf{w}_k} 
    \sum_{j=1}^{n_k} 
    \left( \textbf{w}_k^T \textbf{h}_{k,j} - z_{k,j} \right)^2
    \nonumber\\
    \equiv
    \min_{\textbf{w}_k} \sum_{j=1}^{n_k} 
    \bigg\{
    \left(\textbf{w}_k^T \textbf{h}_{k,j}\right)^2 - 2 \textbf{w}_k^T \textbf{h}_{k,j} z_{k,j}
    \bigg\}
    \nonumber\\
    =
    \min_{\textbf{w}_k}
    \textbf{w}_k^T \textbf{H}_{k} \textbf{H}_{k}^T \textbf{w}_k
    - 2 \textbf{w}_k^T \textbf{H}_{k} \textbf{z}_{k}
\end{align}
where $\textbf{H}_{k} \coloneqq [\textbf{h}_{k,1} ~ \textbf{h}_{k,2} ~ \dots ~ \textbf{h}_{k,n_k}]$ and $ \textbf{z}_k \coloneqq [z_{k,1} ~ \dots ~ z_{k,n_k}]^T$.
The optimization problem in (\ref{eqn: quadratic}) is a quadratic one in the form of $\frac{1}{2}\textbf{w}_k^T \textbf{A}_k \textbf{w}_k + \textbf{w}_k^T \textbf{b}_k$ with $\textbf{A}_k = 2\textbf{H}_{k} \textbf{H}_{k}^T$ and $\textbf{b}_k = - 2 \textbf{H}_{k} \textbf{z}_{k}$.
The solution to this optimization problem when $\textbf{A}_k$ is positive semi definite is $\textbf{w}_k^{\text{opt}} = -\textbf{A}_k^{-1} \textbf{b}_k $ (see e.g. \cite{boyd2004convex}).
However, $\textbf{A}_k$ might become seriously ill conditioned especially when the number of reservoir outputs that had led to action $a_k$ are not plenty enough.
In such a case, we would get extremely high values for the elements of the $\textbf{w}_k$ vector.
Consequently, we will add a regularizer term to the optimization problem in (\ref{eqn: quadratic}) to mitigate this issue.
We therefore try to solve the following problem instead
\begin{align}
    \label{eqn: final regularized}
    &\min_{\textbf{w}_k}
    \textbf{w}_k^T \textbf{H}_{k} \textbf{H}_{k}^T \textbf{w}_k
    - 2 \textbf{w}_k^T \textbf{H}_{k} \textbf{z}_{k} 
    + \frac{1}{2}\alpha \textbf{w}_k^T \textbf{w}_k
    \nonumber\\
    =&\min_{\textbf{w}_k}
    \frac{1}{2}\textbf{w}_k^T \left(2\textbf{H}_{k} \textbf{H}_{k}^T + \alpha \textbf{I} \right) \textbf{w}_k
    + \textbf{w}_k^T \left( -2\textbf{H}_{k} \textbf{z}_{k} \right) 
\end{align}
where $\alpha$ is the regularization parameter.
By setting $\Tilde{\textbf{A}}_k = 2\textbf{H}_{k} \textbf{H}_{k}^T+ \alpha \textbf{I}$ we would have a much more stable solution of $\textbf{w}_k = -\Tilde{\textbf{A}}_k^{-1} \textbf{b}_k $. 

With the necessary building blocks set in place, we can run the RL framework for one episode of $T$ time steps and gather enough samples to form the loss minimization in (\ref{eqn: loss decomposition}).
We then find the minimum value of the bias term through finding the optimum weights of each of the rows of the output weight matrix $W_{\text{out}}$ as was outlined in (\ref{eqn: final regularized}).
We repeat this for multiple choices of the recurrent and input weight matrices and compare their bias plus variance error and check to see which one has the least error.
We then choose that ESN as our main Q network and go forward with the RL training in the next episodes using Algorithm \ref{alg:DEQN_training}.

%

%
\section{RL Problem Formulation for 5G DSA}
\label{sec:problem_definition}

\label{sec:DSA_formulation}
In this section, we first outline two challenges that we face in solving real world problems using RL. 
We then introduce the DSA problem and discuss the setting that we choose for our RL formulation. 

\subsection{Motivations}
In many real-world scenarios, we encounter environments with a highly dynamic behaviour, where the agent has to deal with rewards that change behavior over time and variable transition probabilities between system states through time.
These kinds of environments degrade the performance of DRL-based strategies drastically because the learned policy is not suitable for the latest environment.
Furthermore, the high variance poses a big challenge to learning stability because the collected training data become obsolete as the environment changes~\cite{foerster2017Replay}, which means the effective training data is limited. 
Therefore, we adopt an online training algorithm that only relies on the very recent experiences and removes outdated training data from the replay buffer to update the policy, so that it can adapt to the environment quickly.
%
%
Most existing DRL algorithms require a large number of samples for the training to converge.
This may not be an issue for applications such as computer games where samples can be easily obtained.
However, in 5G networks, data collection is usually expensive making it challenging to directly apply DRL algorithms.
Furthermore, the 5G environment is constantly changing where the collected past data become obsolete and thus there is very limited valid training data available.
As a result, developing efficient DRL algorithms requiring limited training data with fast convergence is critical to make DRL applicable to 5G networks.
With these motivations in mind, we will adopt the DEQN to tackle the DSA problem.
We will also use the method outlined in section~\ref{sec:theoretical} to determine the best  hyperparameter $B_{rec}$ in our DEQN framework.

\subsection{DSA Problem formulation }
DSA is a promising technology to improve the utilization of radio spectrum. 
DSA allows an SU to access the radio spectrum if PUs only receive tolerable interference. 
The primary system consists of a primary base station (PBS) and $M$ PUs, and the secondary system consists of a secondary base station (SBS) and one or more SUs.
The primary system has a license to operate on $C$ wireless channels.
It is assumed that each user can only transmit on one channel at a particular time.
Users cause interference to one another if they transmit on the same channel at a particular time.
Our goal is to develop a spectrum access strategy of the SU to increase the overall spectrum utilization without generating intolerable interference to PUs.
%
%

%
A spectrum opportunity occurs on a channel for two cases: 1) There is no PU that conducts data transmission on that channel. 2) The SU can share a channel with the PU if the interference to the PU is tolerable.
Unfortunately, obtaining this control information is costly in 5G mobile wireless networks.
Furthermore, the control information is outdated quickly in the non-stationary 5G networks.
Therefore, it is impractical to design a DSA strategy by assuming that all the control information is known.

The SU causes interference to PUs if they conduct data transmission on the same channel at the same time.
To protect the primary system, the SU performs spectrum sensing to detect the activity of PUs before accessing a channel.
%
We assume that the SU has low communication capabilities and can only sense one channel at a particular time.
There has been studies on definition of reduced capability devices in ~\cite{reduced-capability}.
The DRL agent will use the sensing outcomes as the input to learn a spectrum access strategy in order to maximize the spectrum utilization while enabling the protection to PUs.
To enhance the protection to the primary system, we assume that the primary system will broadcast a warning signal if it experiences strong interference.
The warning signal contains information related to which PU experiences interference so that the SU is aware of the issue. 

\subsection{DRL-based solution for DSA}
We now formulate the DSA problem using the DRL framework. 
To be specific, we assume that the SU has a DRL agent that takes the sensing outcomes as observations and learns how to perform spectrum sensing and access actions in order to maximize its cumulative reward. 
Since the spectrum sensing outcomes are incomplete and noisy, the environment is partially observable.
Suppose that there are a total of $C$ channels available for transmission.
The observation of the SU at time $t$ is denoted by $s_t = \left(c_t, q_t \right)$, where $c_t \in \{0,1\}$ is the sensing outcome at time $t$ and $q_t$ is a one-hot $C$-dimensional vector representing what channel was sensed at time $t$.
We assume that the SU performs spectrum sensing and then it makes the spectrum access decision.
The action of the SU at time $t$ is denoted by $a_t = \left( u_t, i_t \right)$, where $u_t \in \{0, 1\}$ represents whether the SU will access the current sensed channel ($u_t = 1$) or stay idle ($u_t = 0$) at time $t$, $i_t \in \{1, ..., C\}$ represents the channel index that the SU will sense at time $t+1$.
In other words, the SU makes two decisions: $u_t$ decides whether to conduct data transmission in the current sensed channel at time $t$ and $i_t$ decides which channel to sense at time $t+1$.

The reward for the SU is designed to maximize the spectrum utilization while preventing harmful interference to PUs, which is based on the underlying achieved modulation and coding (MCS) strategy adopted in the 3GPP LTE/LTE-Advanced standard~\cite{LTE}.
To be specific, a user measures the quality of the wireless connection and determines the MCS for data transmission.
The selected MCS corresponds to a specific spectral-efficiency, which is defined as the transmitted information bits per symbol in a range from $0.1523$ to $5.5547$.
The reward function depends on the average spectral-efficiency of PUs, and we choose $3$ as the threshold for determining if the interference to the primary system is tolerable.
Let $\bar{e}_{c, t}^{\text{PU}}$ be the average spectral-efficiency of PUs on channel $c$.
The reward of the SU at time $t$ is written as
\begin{equation}
{r_t} = 
\begin{cases}
0, & \text{SU is idle},\\
-1, & \text{SU accesses channel $c$ and } \bar{e}_{c, t}^{\text{PU}} < 3, \\
1, & \text{SU accesses channel $c$ and } \bar{e}_{c, t}^{\text{PU}} \geq 3.
\end{cases}
\label{eqn:reward_def}
\end{equation}
With the state space, action space and the reward function defined, we are able to set an SU as an RL agent and let it learn from the environment the best actions to take.
In the next section we will provide simulation results of applying the mentioned RL framework.

\section{Performance Evaluation}
\label{sec:performance}
In this section, we evaluate the performance of the introduced DEQN methodology in the DSA scenario of a relevant 5G network.
We develop a network simulator to provide practical radio environments in 5G networks by incorporating field measurements obtained from a $10$-cell ray-tracing area in a city and real-world spectrum occupancy database~\cite{wellens2010lessons}.
Furthermore, all users are moving based on the random waypoint model~\cite{hyytia2007rwp}, which is commonly used for modeling the user mobility in 5G networks.
%
%
%
%
%
%


\begin{figure}[t]
\centering
\begin{subfigure}[b]{0.48\linewidth}
\includegraphics[width=1\linewidth]{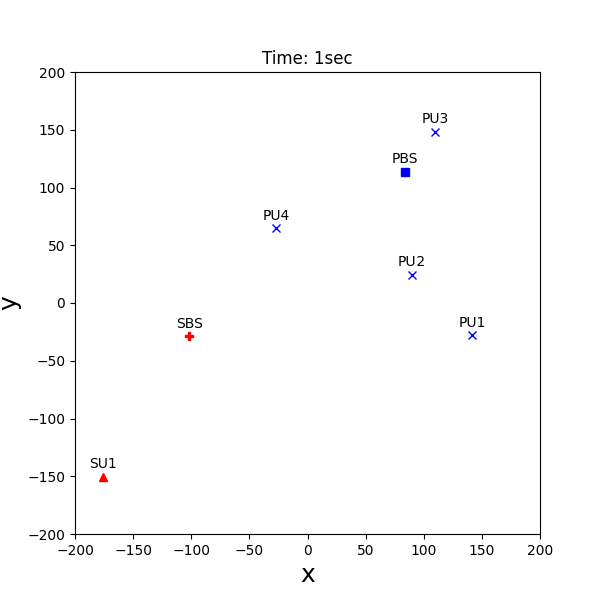}
\caption{}
\end{subfigure}
\begin{subfigure}[b]{0.48\linewidth}
\includegraphics[width=1\linewidth]{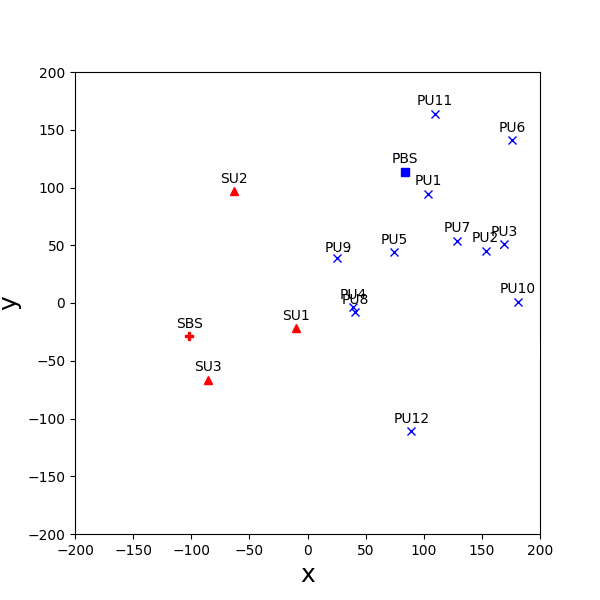}
\caption{}
\end{subfigure}
\caption{A snapshot of the DSS network geometry at the beginning of the simulation. (a) First simulation scenario with 4 PUs and 1 SU. (b) Second simulation scenario with 12 PUs and 3 SUs.}

\label{fig:geometry} 

\end{figure}

Fig. \ref{fig:geometry} shows the DSA scenarios in our experiment, where there are $1$ SU and $4$ PUs in scenario (a), and $3$ SUs and $12$ PUs in scenario (b) for a 400m$\times$400m area.
The primary system has a license to operate on $3$ wireless channels ($d_x=4$ and $d_y=6$), while each SU has to access the channels without generating intolerable interference to the primary system.
Each SU adopts an online training algorithm that learns its spectrum access strategy continually to adapt to the non-stationary 5G environment.
To be specific, each SU collects $T = 500$ training samples in one episode and stores them in its memory buffer.
Then the SU trains its DRL agent for $100$ iterations to update its policy and applies the updated policy in the next episode.
Since the environment gradually changes over time, the SU removes the outdated training samples from the memory buffer and collects another $T$ training samples in the new episode.
%

%
%
Clearly, the second scenario is a more dynamic environment and due to the presence of multiple agents, it becomes even more dynamic as the agents also change their behavior over time.
%
We analyze the effectiveness of our theoretical framework in scenario (a), because there is only one agent in the environment and the underlying assumption of the system having the Markov property is not voided.
Specifically, we apply the method outlined in section \ref{subsection:empirical-bias} to find the best $B_{rec}$ to use for our ESN kernel of the Q-network.
We then check if the suggested $B_{rec}$ provides good performance for us.
In the second scenario, however, we aim to evaluate the effect of using ESN as the recurrent Q-network, in comparison with two baselines that use different recurrent Q-network structures, which are DRQN \cite{hausknecht2015DRQN} and ADRQN \cite{zhu2018improving}.
ADRQN extends DRQN by using both observation and action as inputs of the Q-network.
Although we assume vanilla RNN in theoretical analysis for DRQN to simplify the analysis, Long Short Term Memory (LSTM) is usually preferred to be used in DRQN.
Therefore, LSTM is used as the Q-network for DRQN and ADRQN.
To have a fair comparison, the LSTM of DRQN/ADRQN and the ESN of DEQN have the same number of neurons $d_h = 64$ and the learning rate that we saw has the best performance on each kernel.
This was $0.01$ for the DEQN and $0.001$ for the DRQN and ADRQN.
Furthermore, we let $B_X = 1$, $B_{\text{in}} = 0.5$, $B_{\text{out}}=10$, $\gamma = 0.9$, $R_{\text{max}} = 1$, and $\epsilon = 1.5$.
The choice of $\epsilon$ should usually be around $1/10$ of $V_{max} = R_{max}/(1-\gamma)$.
%

\begin{figure}[t]
\centering
\includegraphics[width=.7\linewidth]{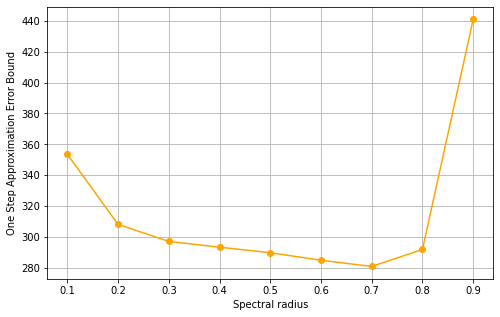}
	\caption{The bound on the one step approximation error  with respect to ESN's recurrent weights spectral radius, for the case of 4 PUs and 1 SU.}
	\label{fig:recurrent_norm}

\end{figure}

\begin{figure}[t]
\centering
	\includegraphics[width=.85\linewidth]{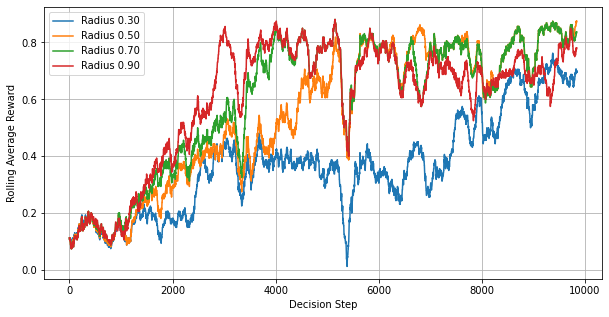}
	\caption{Rolling average reward under different recurrent weights' spectral radius per each step of decision making in the case of 4 PUs and 1 SU.}
	\label{fig:mean_reward_recurrent_norm}


\end{figure}

From the method outlined in section \ref{subsection:empirical-bias}, we calculate the sum of the squared bias and the variance in scenario (a) under different $B_{\text{rec}}$ in Fig. \ref{fig:recurrent_norm}. 
We observe that $B_{\text{rec}} = 0.7$ achieves the lowest upper bound on one step approximation error.
Fig. \ref{fig:mean_reward_recurrent_norm} shows the rolling window average reward achieved under different choices of $B_{\text{rec}}$ with a window length of 160 decision steps.
We can see that the choice of $B_{\text{rec}}=0.7$ ultimately achieves better performances than other choices while maintaining a good convergence rate.
We similarly set the $B_{rec}$ to be $0.7$ in scenario (b) too.
%



\begin{figure}[t]
\centering
\includegraphics[width=0.88\linewidth]{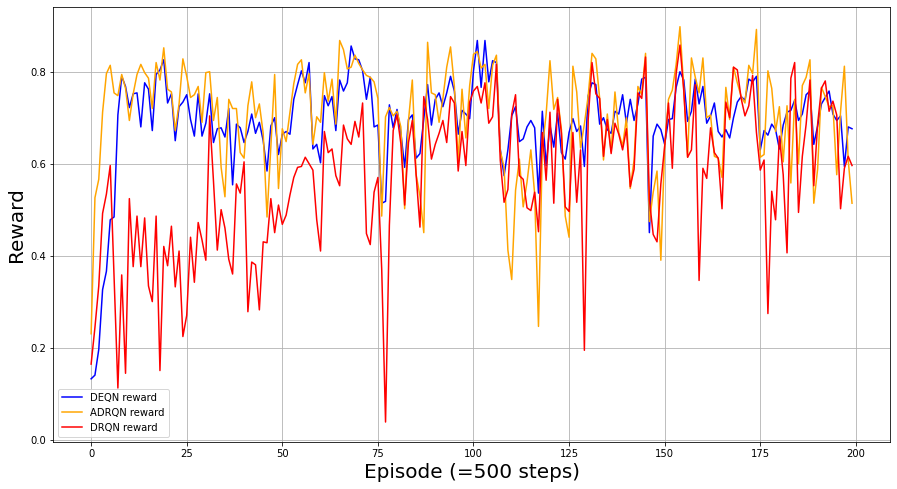}
\caption{Average reward per episode with 4 PUs and 1 SU.}
\label{fig:result_reward_1SU}

\end{figure}

\begin{figure}[t]
\centering
\includegraphics[width=0.88\linewidth]{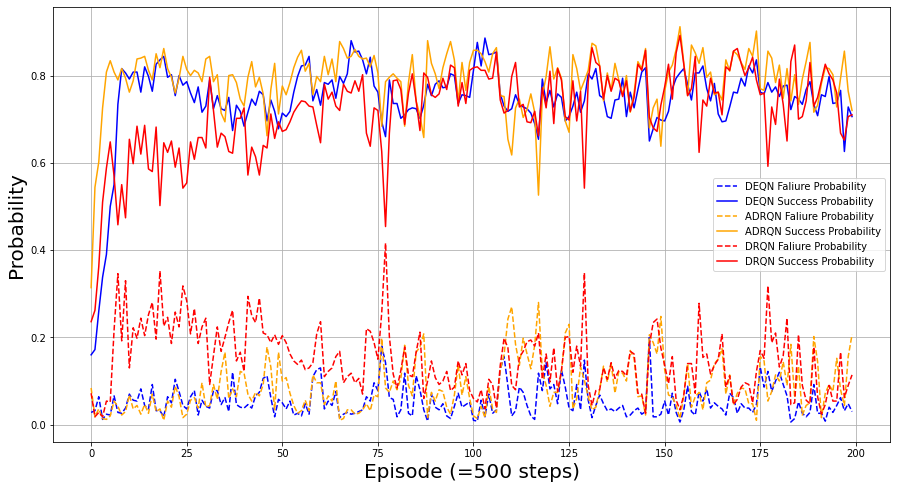}
\caption{Simulation results with 4 PUs and 1 SU. (a) Average reward per episode. (b) Success and failure probability per episode for accessing a channel.}
\label{fig:result_success_prob_1SU}

\end{figure}

We now do the comparison between DEQN and DRQN in terms of their performance in a DSA environment.
%
%
%
Specifically, we will compare the average received reward per episode and the probability of success and failure in accessing a channel.
All the results are compared while randomly initializing the networks. Each episode is considered to be 500 steps of decision making.
%

The curves of the average reward per episode and success/failure probability is shown in Figs. (\ref{fig:result_reward_1SU}) and (\ref{fig:result_success_prob_1SU}) respectively for each Q-network kernel.
From both of these figures we observe that DEQN and ADRQN have roughly the same convergence rate while DRQN needs many more training samples to reach its best performance.
The comparable performance of DEQN with ADRQN is while DEQN has a simple vanilla RNN type structure and the hidden state variables are not even trained, compared to the other two methods that use LSTM with trainable parameters.
Assuming that the device has low computational capabilities, it is important to choose a learning method that has less complexity.
Therefore, we will also compare the total simulation time of different learning methods.
Table \ref{tab:training_time_1SU} compares the average total CPU run time for $200$ episodes using the same machine (2.71 GHz Intel i5 CPU) in scenario (a).
DEQN has the shortest training time because it only trains the output weights and avoids recalculating hidden states. 
This is also an important aspect of DEQNs, as we need computationally inexpensive algorithms that can be used in regular low cost user equipment.

\begin{figure}[t]
\centering
\includegraphics[width=1\linewidth]{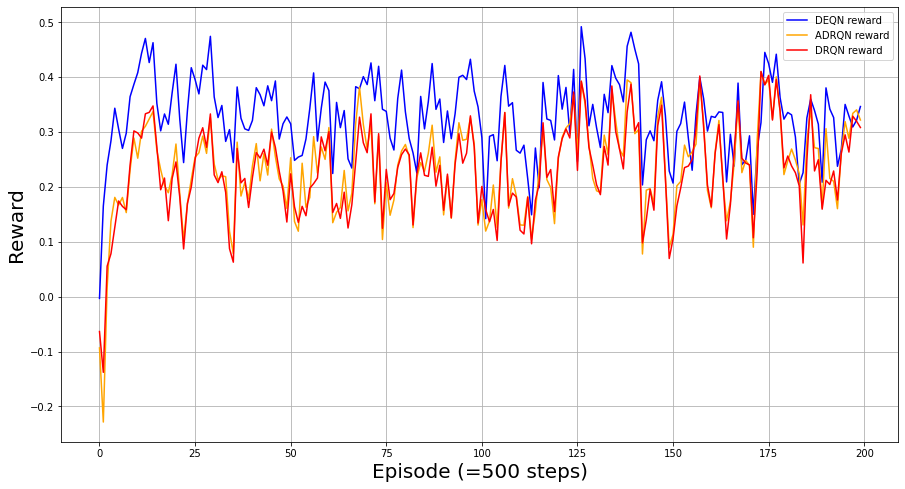}
\caption{Average reward per episode with 12 PUs and 3 SUs. Every plot is an average over secondary users.}
\label{fig:result_reward}

\end{figure}

\begin{figure}[t]
\centering
\includegraphics[width=0.88\linewidth]{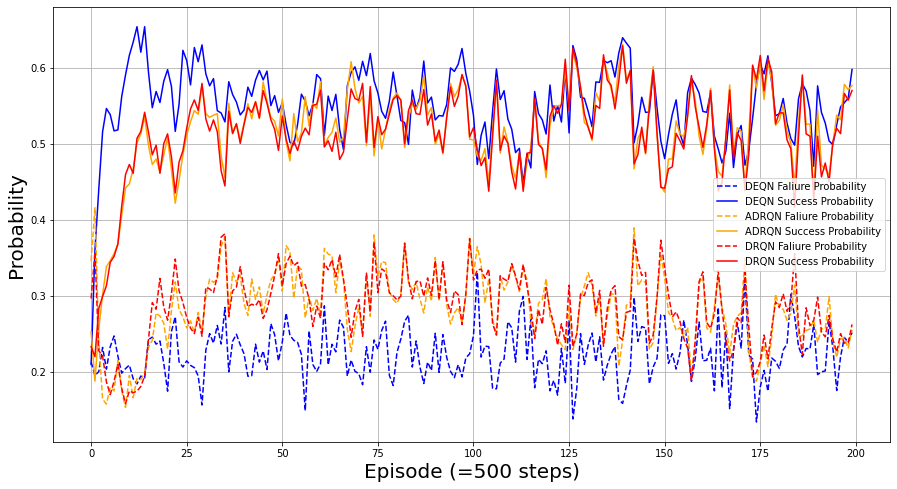}
\caption{Success and failure probability per episode for accessing a channel with 12 PUs and 3 SUs. Every plot is an average over secondary users.}
\label{fig:result_success_prob}

\end{figure}

\begin{table}[ht]
	\centering
	\caption{Average CPU run time with 4PUs and 1 SU.}
	\label{tab:training_time_1SU}  
	\begin{tabular}{|c|c|c|}
	\hline
	DEQN & DRQN & ADRQN \\ \hline
	$330.4$sec  & $1255.9$sec & $1417.5$sec\\ \hline
    \end{tabular}
\end{table}
We emphasize that the the key benefit of using ESNs as kernel is its higher convergence rate.
This feature is mostly prominent in highly varying and dynamic environments.
Therefore, we will showcase the results for the case of scenario (b) too.
Similarly, Figs. (\ref{fig:result_reward}) and (\ref{fig:result_success_prob}) show the average reward and success/failure probability per episode respectively.
We can see that DEQN reaches it's best performance after around 25 episodes, while the other two methods will need more than 150 episodes of training to reach their best performance.
Note that the most received reward for every DQN structure has decreased compared to scenario (a). This is because the environment has more PUs that are constantly transmitting data and the agents seem to have decided to stay idle more than the case in scenario (a). On the other hand, the success rate of DRQN and ADRQN reaches its best relatively quickly, while it takes a large number of samples for them to reduce their failure rate. 
This is while DEQN has a much faster convergence rate and has already learned to lower its failure rate while maintaining a high success rate.
Therefore, DEQN can find the most spectrum opportunities.
Similarly to scenario (a), we compare the average total CPU run time in scenario (b) (summed over all agents) in Table \ref{tab:training_time}. It is apparent that DEQN takes much less time to train its network weights than DRQN. 
%

%

%



Since all methods have the same number of training samples and training iterations, the evaluation results confirm the theoretical findings and clearly demonstrate that DEQN has the benefit of faster learning under limited training data than DRQN in practice.

\begin{table}[ht]
	\centering
	\caption{Average CPU run time with 12 PUs and 3 SUs}
	\label{tab:training_time}  
	\begin{tabular}{|c|c|c|}
	\hline
	DEQN & DRQN & ADRQN \\ \hline
	$630.5$sec  & $2806.4$sec & $2911.4$sec\\ \hline
    \end{tabular}
\end{table}

\section{Conclusion}
\label{sec:conclusion}
In this work, we studied the performance of DRQNs and DEQNs and compared their performance using the mathematical notion of covering numbers and how it affects the one step approximation error of a DQN. 
We also proposed a method for determining the hyper parameters of a DEQN by minimizing the empirically derived bias error plus the theoretically upper bounded variance error.
We showed that DEQNs can provide faster convergence rates due to a lower covering number bound.
Our simulation of a dynamic spectrum sharing network with multiple SUs and PUs also verified this as there were multiple agents in the environment, adding to its non-stationarity.
It is absolutely critical to use a framework that converges faster in these environments as the training data becomes outdated quickly. 
DEQNs provide us such a framework, allowing us to not only train our network with less training data, but also incur low computational compelxity for the user equipment.


\begin{appendices}
\section{Proof of RNN covering number bound}
\label{apx: RNNproof}
In order to move forward with the proof, we first need to introduce a Lemma from~\cite{Bartlett2017} which provides an upper bound for linear predictors handled with matrix multiplications.
\begin{lemma}
    \label{lmma: Bartlett}
    (Lemma 3.2 in~\cite{Bartlett2017}) Let $X\in \mathbb{R}^{d\times n}$ be a given data matrix with $\|X\|_F \le b$,
    where $d$ is the number of features and $n$ is the number of samples.
    Then
    \begin{align}
        \ln \mathcal{N}
        \left( \Big\{
            WX : W\in \mathbb{R}^{m\times d}, \|W\|_F \le a
        \Big\}, \epsilon, \|\cdot\|_2 \right)
        \nonumber\\
        \le
        \left( \dfrac{a^2 b^2 m}{\epsilon^2} \right)
        \ln (2dm)
    \end{align}
\end{lemma}
In what follows, we first find a cover set for the whole network by introducing a cover for each matrix multiplication in the RNN.
Then, we will tune each of the introduced covers' granularity such that the whole cover would have the desired granularity. 
\subsection{Covering the whole network}
We will start by forming a cover set for each of the inputs, i.e. $X_i , 1\le i\le t$.
For each input $X_i$, we form an $\varepsilon_i$-cover set $\mathscr{X}_i$ of $W X_i$.
The cardinality of $\mathscr{X}_i$ is
\begin{align}
    |\mathscr{X}_i|
    &=
    \mathcal{N}\Big\{ W X_i: W\in \mathbb{R}^{d_h \times d_x} , \|W\|_F \le B_{in},\varepsilon_i,\|\cdot \|_2 \Big\}
    \nonumber\\
    &\le
    N_i
    \label{inputLayerCover}
\end{align}
where $N_i$ is obtained from the RHS of Lemma~\ref{lmma: Bartlett}. 

We then inductively form cover sets for each of the time slots of RNN.
For each element $F\in \mathscr{X}_i$, we form a $\mu_1$ cover $\mathscr{U}_1(F)$ of $U(\sigma_h(F))$. The cardinality of $\mathscr{U}_1(F)$ is
\begin{align}
    |\mathscr{U}_1(F)|=\mathcal{N}\Big\{& U(\sigma_h(F)): U\in \mathbb{R}^{d_h \times d_h} ,
    \|U\|_F \le B_{rec},
    \mu_1,
    \nonumber\\
    &\|\cdot \|_2 \Big\} \le M_1 ; F\in \mathscr{X}_1
    \label{firstRecurrentLayerCover}
\end{align}
where $M_1$ is obtained from the RHS of Lemma~\ref{lmma: Bartlett}. We then form the cover $\mathscr{H}_1 \coloneqq \bigcup_{F\in \mathscr{X}_i} \mathscr{U}_1(F)$ whose cardinality is
\begin{align}
    \big | \mathscr{H}_1 \big | = \Bigg | \bigcup_{F\in \mathscr{X}_i} \mathscr{U}_1(F)\Bigg | \le \sum_{F\in \mathscr{X}_i} \big | \mathscr{U}_1(F) \big | \le M_1 N_1
    \label{FirstH}
\end{align}
Next, for every cover element $F\in \mathscr{X}_i$ and every cover element $\Tilde{F}\in \mathscr{H}_{i-1}$, we form a $\mu_i$ cover $\mathscr{U}_i(F,\Tilde{F})$ of $U(\sigma_h(F+\Tilde{F}))$. The cardinality of $\mathscr{U}_i(F,\Tilde{F})$ is
\begin{align}
    &|\mathscr{U}_i(F,\Tilde{F})|=\mathcal{N} \Big\{ U(\sigma_h(F+\Tilde{F})):
    U\in \mathbb{R}^{d_h \times d_h},
    \|U\|_F \le B_{rec},
    \nonumber\\
    &\mu_i,\|\cdot \|_2 \Big\}
    \le M_i ;
    i\in \{2,\dots,t\} ,F\in \mathscr{X}_i, \Tilde{F}\in \mathscr{H}_{i-1}
    \label{recurrentLayersCovers}
\end{align}
where $M_i$ is obtained from the RHS of Lemma~\ref{lmma: Bartlett}. We then form the cover $\mathscr{H}_i \coloneqq \bigcup_{F\in \mathscr{X}_i, \Tilde{F}\in \mathscr{H}_{i-1}} \mathscr{U}_i(F,\Tilde{F})$ whose cardinality is
\begin{align}
    \big | \mathscr{H}_i \big | = \Bigg | \bigcup_{\substack{F\in \mathscr{X}_i\\ \Tilde{F}\in \mathscr{H}_{i-1}}} , \mathscr{U}_i(F, \Tilde{F}) \Bigg | 
    &\le
    \sum_{F\in \mathscr{X}_i} \sum_{\Tilde{F}\in \mathscr{H}_{i-1}} \big | \mathscr{U}_1(F , \Tilde{F}) \big | 
    \nonumber\\ 
    &\le
    M_i N_i \big | \mathscr{H}_{i-1} \big |
\label{InductiveCoverBoundH}
\end{align}
Therefore, by an induction on (\ref{InductiveCoverBoundH}) given the first expression as (\ref{FirstH}), we would have
\begin{align}
    \big | \mathscr{H}_i \big |
    \le
    \prod_{k=1}^i M_k N_k
\end{align}
Finally, for every cover element $F\in \mathscr{X}_t$ and every cover element $\Tilde{F} \in \mathscr{H}_{t-1}$ we form the $\delta$-cover set $\mathscr{Y}_t (F,\Tilde{F}$) of $V(\sigma_h(F+\Tilde{F}))$.
The cardinality of $\mathscr{Y}_t (F,\Tilde{F})$ is
\begin{align}
    &|\mathscr{Y}_t(F,\Tilde{F})|
    =
    \mathcal{N} \Big\{ V(\sigma_h(F+\Tilde{F})): V\in \mathbb{R}^{d_y \times d_h},
    \|V\|_F \le B_{out},
    \nonumber\\
    &\delta,
    \|\cdot \|_2 \Big\}
    \le Q ;
    F\in \mathscr{X}_t, \Tilde{F}\in \mathscr{H}_{t-1}
    \label{outputLayerCover}
\end{align}
where $Q$ is obtained from the RHS of Lemma~\ref{lmma: Bartlett}. We then form the whole cover of the RNN function class by taking a union over $F$ and $\Tilde{F}$ in (\ref{outputLayerCover}),
i.e. $\mathscr{F}_t \coloneqq \bigcup_{F\in \mathscr{X}_t, \Tilde{F}\in \mathscr{H}_{t-1}} \mathscr{Y}_t(F,\Tilde{F})$. The cardinality of $\mathscr{F}_t$ is lower bounded by
\begin{align}
    \big | \mathscr{F}_t \big | = \Bigg | \bigcup_{\substack{F\in \mathscr{X}_t\\ \Tilde{F}\in \mathscr{H}_{t-1}}} \mathscr{Y}_t(F,\Tilde{F})\Bigg | 
    &\le
    \sum_{F\in \mathscr{X}_t} \sum_{\Tilde{F}\in \mathscr{H}_{t-1}} \big | \mathscr{Y}_t(F,\Tilde{F}) \big | 
    \nonumber\\ 
    &\le
    N_t \Bigg( \prod_{k=1}^{t-1} M_k N_k \Bigg ) Q
    \label{wholeCover}
\end{align}

\subsection{Tuning the spacing between the cover elements}
In this section we will intentionally set $\varepsilon_i$'s, $\mu_i$'s and $\delta$ in such a way that $\mathscr{F}_t$ would be an $\varepsilon$-cover for $Y_t$.
In other words, we need those variables to be set in such a way that for any given set of network parameters, we would be able to present a $\hat{Y}_t\in \mathscr{F}_t$ such that $|Y_t - \hat{Y}_t| \le \varepsilon$. 

To proceed, we will fix arbitrary values for $U$, $W$ and $V$.
This way, we would fix a function in our function class and we would like to find a $\hat{Y}_t$ which is close to the output of this function given a specific set of inputs $\{ X_1, X_2, \dots, X_t \}$.
We recursively define the outputs of each time-series layer to be
\begin{align}
    &F_i \coloneqq W X_i \nonumber\\
    &G_0 \coloneqq 0 \nonumber\\
    &G_i \coloneqq U(\sigma_h(F_i + G_{i-1})) \nonumber \\
    &J_t \coloneqq V(\sigma_h(F_i + G_{i-1}))
\end{align}
Next, we choose a cover element close to the output of each input layer. Then, we recursively choose cover elements close the output of each time-series layer, given the cover elements of the previous layer as inputs. The choices are as follows
\begin{itemize}
    \item Choose $\hat{F}_i\in \mathscr{X}_i$ such that $\| W X_i - \hat{F}_i \| < \varepsilon_i$
    \item Choose $\hat{G}_i\in \mathscr{H}_i$ such that $\| U(\sigma_h(\hat{F}_i + \hat{G}_{i-1})) - \hat{G}_i \| < \mu_i$
    \item Choose $\hat{J}_t\in \mathscr{Y}_t$ such that $\| V(\sigma_h(\hat{F}_t + \hat{G}_{t-1})) - \hat{J}_t \| < \delta$
    \item Finally, let $\hat{Y}_t = \sigma_y (\hat{J}_t)$
\end{itemize}

Now that we have properly chosen our cover elements, we need to make sure the cover element chosen for the output is close enough to the actual output. Therefore, we will write
\begin{align}
    &|Y_t - \hat{Y}_t|
    \le
    \rho_y |J_t - \hat{J}_t| \nonumber\\
    &\le
    \rho_y \Big|V\big(\sigma_h(U H_{t-1}+W X_t )\big) - V\big(\sigma_h(\hat{G}_{t-1} + \hat{F}_t)\big) \nonumber \\
    &+ V\big(\sigma_h(\hat{G}_{t-1} + \hat{F}_t)\big) - \hat{J}_t\Big| \nonumber \\
    & \le
    \rho_y \Big( |V|_{op}\rho_h \big| U H_{t-1} + W X_t - \hat{G}_{t-1} - \hat{F}_t \big| + \delta \Big) \nonumber \\
    &\le
    \rho_y\rho_h |V|_{op} \Big( \big| U H_{t-1} - \hat{G}_{t-1} \big| + \varepsilon_t \Big) + \rho_y \delta \nonumber \\
    & =
    \rho_y \delta + \rho_y \rho_h |V|_{op} \varepsilon_t + \rho_y \rho_h |V|_{op} \big| U H_{t-1} - \hat{G}_{t-1} \big|
    \label{outputDifference}
\end{align}
The first two terms in (\ref{outputDifference}) can be controlled using $\delta$ and $\varepsilon_t$. In order to control the last term, however, we will conduct the following analysis
\begin{align}
    &\big| U H_{i} - \hat{G}_{i} \big|
    =
    \Big| U H_{i} -U\big(\sigma_h(\hat{G}_{i-1} + \hat{F}_i)\big) \nonumber \\
    &+ U\big(\sigma_h(\hat{G}_{i-1} + \hat{F}_i)\big) - \hat{G}_{i} \Big| \nonumber \\
    &\le
    |U|_{op} \Big| H_i - \sigma_h \big( \hat{G}_{i-1} + \hat{F}_i \big) \Big| + \mu_i \nonumber \\
    &=
    |U|_{op} \Big| \sigma_h\big( U H_{i-1} + W X_{i} \big) - \sigma_h \big( \hat{G}_{i-1} + \hat{F}_i \big) \Big| + \mu_i \nonumber \\
    & \le
    \rho_h |U|_{op} \big| U H_{i-1} - \hat{G}_{i-1} + W X_{i} - \hat{F}_i \big| +\mu_i \nonumber \\
    & \le
    \rho_h |U|_{op} \big| U H_{i-1} - \hat{G}_{i-1}\big| +
    \rho_h |U|_{op} \varepsilon_i + \mu_i
    \label{inductiveDistanceH}
\end{align}
Therefore, by an induction on (\ref{inductiveDistanceH}) while considering the fact that $\big| U H_0 - \hat{G}_0 \big|=0$, we would have
\begin{align}
    \big| U H_{t-1} - \hat{G}_{t-1} \big|
    \le
    \sum_{k=1}^{t-1} {\rho_h^k |U|_{op}^k \varepsilon_{t-k}}
    +\sum_{k=0}^{t-2}\rho_h^k |U|_{op}^k \mu_{t-1-k}
    \label{StateDifference}
\end{align}
Inserting (\ref{StateDifference}) into (\ref{outputDifference})
and considering the fact that for every matrix $A$, $\|A\|_{op} \le \|A\|_F$,
we get
\begin{align}
    &|Y_t - \hat{Y}_t|
    \le 
    \rho_y \delta 
    + \rho_y \rho_h |V|_{F} \varepsilon_t 
    + \rho_y \rho_h |V|_{F} \sum_{k=1}^{t-1} {\rho_h^k |U|_{F}^k \varepsilon_{t-k}} \nonumber\\
    &+ \rho_y \rho_h |V|_{F} \sum_{k=0}^{t-2}\rho_h^k |U|_{F}^k \mu_{t-1-k}
    \nonumber\\
    &\le 
    \rho_y \delta 
    + \rho_y \rho_h B_{out} \left( \sum_{k=0}^{t-1} {\rho_h^k B_{rec}^k \varepsilon_{t-k}} 
    + \sum_{k=0}^{t-2}\rho_h^k B_{rec}^k \mu_{t-1-k} \right)
\end{align}
Now we have reached a point where we can fully control how close the cover element in $\mathscr{F}_t$ can be to $Y_t$. We simply set
\begin{align}
    \label{eqn: delta}
    &\delta = \frac{\varepsilon/3}{\rho_y} \\
    \label{eqn: epsilon_i}
    & \varepsilon_i = \frac{\varepsilon/3}{t \rho_y \rho_h B_{out} \rho_h^{t-i}B_{rec}^{t-i} }, i\in \{1,\dots,t \} \\
    \label{eqn: mu_i}
    &\mu_i = \frac{\varepsilon/3}{(t-1)\rho_y \rho_h B_{out} \rho_h^{t-1-i}B_{rec}^{t-1-i} }, i\in \{1,\dots,t-1 \}
\end{align}
so that we would have
\begin{align}
    |Y_t - \hat{Y}_t| \le \varepsilon
\end{align}

Now that we have covered the output of an RNN, it is required that we characterize $N_i$, $M_i$ and $Q$ so that we could evaluate (\ref{wholeCover}).
For this end, we need to get an upper bound on the norm of the output of each layer.

Starting with (\ref{inputLayerCover}), we have $\|X\|_F \le B_X$.
For the coverings in (\ref{firstRecurrentLayerCover}), (\ref{recurrentLayersCovers}) and (\ref{outputLayerCover}), the input of the covers are $\sigma_h (H_i) = \sigma_h (WX_i + UH_{i-1})$, which is spectrally bounded by
\begin{align}
    &\|\sigma_h (WX_i + UH_{i-1})\|
    \le
    \rho_h(\|WX_i\|+\|UH_{i-1}\|) \nonumber\\
    &\le
    \rho_h\|W\|\|X_i\| + 
    \rho_h\|U\|\|H_{i-1}\| \nonumber \\
    &=\rho_h\|W\|\|X_i\| + 
    \rho_h\|U\|\|\sigma_h (WX_{i-1}+UH_{i-2})\|,  \nonumber\\
    &i\in \{1,\dots,t\},X_0=H_0=H_{-1}=0
\end{align}
We can then use induction to conclude that
\begin{align}
    &\|\sigma_h (H_i)\|
    =\|\sigma_h (WX_i + UH_{i-1})\|\nonumber\\
    &\le
    \sum_{k=1}^i {\rho_h^{i-k+1}}\|W\|\|X_k\| \|U\|^{i-k}
    \le
    B_W B_X \sum_{k=1}^i {\rho_h^{i-k+1}} B_{rec}^{i-k} \nonumber\\
    &=
    \rho_h B_W B_X \dfrac{\rho_h^i B_{rec}^i -1}{\rho_h B_{rec} -1}
    \eqqcolon B_{H,i}
    ,i\in \{1,\dots,t\}
    \label{HiUpperBound}
\end{align}
Therefore, we would have
\begin{align}
    \label{eqn: N_i}
    \ln (N_i) 
    \coloneqq
    \left( \dfrac{B_{in}^2 B_X^2 d_h}{\epsilon_i^2} \right)
        \ln (2d_h d_x)
    \\
    \label{eqn: M_i}
    \ln(M_i)
    \coloneqq
    \left( \dfrac{B_{rec}^2 B_{H,i}^2 d_h}{\mu_i^2} \right)
        \ln (2d_h^2)
    \\
    \label{eqn: Q}
    \ln(Q)
    \coloneqq
    \left( \dfrac{B_{out}^2 B_{H,t}^2 d_y}{\delta^2} \right)
        \ln (2d_y d_h)
\end{align}

Combining (\ref{eqn: delta} - \ref{eqn: mu_i}) and (\ref{HiUpperBound}) with (\ref{eqn: N_i} - \ref{eqn: Q}) and substituting in (\ref{wholeCover}) we get
\begin{align}
    \ln \left( |\mathscr{F_t}| \right)
    =
    \ln(Q) + \sum_{k=1}^t \ln(N_i) + \sum_{k=1}^{t-1} \ln(M_k)
\end{align}
in which we have
\begin{align}
    \ln(Q) = \left( \dfrac{9\rho_y^2 \rho_h^2 B_{out}^2 B_W^2 B_X^2 \left( \rho_h^t B_{rec}^t -1 \right)^2}{\left( \rho_h B_{rec} -1\right)^2 \epsilon^2} \right)
    \ln(2d_y d_h)
\end{align}
\begin{align}
    &\sum_{k=1}^t \ln(N_i) 
    =
    \sum_{k=0}^{t-1} \ln(N_{t-k})
    \nonumber\\
    &=
    \sum_{k=0}^{t-1} \dfrac{9t^2 \rho_y^2 \rho_h^2 B_{out}^2 B_{in}^2B_X^2 B_{rec}^2 \rho_h^{2k} }{\epsilon^2} \ln(2d_h d_x)
    \nonumber\\
    &= \dfrac{9t^2 \rho_y^2 \rho_h^2 B_{out}^2 B_{in}^2B_X^2 B_{rec}^2 \rho_h^{2} }{\epsilon^2}
    \left( \dfrac{\rho_h^{2t}B_{rec}^{2t} -1}{\rho_h^{2}B_{rec}^{2} -1} \right)
    \ln(2d_h d_x)
\end{align}
\begin{align}
    &\sum_{k=1}^{t-1} \ln(M_i) 
    =
    \sum_{k=1}^{t-1} \ln(M_{t-k})
    \nonumber\\
    & = \sum_{k=1}^{t-1} \dfrac{9(t-1)^2 \rho_y^2 \rho_h^{2k} B_{out}^2 B_{rec}^{2k} B_{H,t-k}^2 d_h}{\epsilon^2} \ln(2d_h^2)
    \nonumber\\
    & = \dfrac{9(t-1)^2 \rho_y^2 \rho_h^2 B_W^2 B_X^2 B_{out}^2 }{ \epsilon^2 \left( \rho_h B_{rec} -1 \right)^2} 
    \sum_{k=1}^{t-1} \rho_h^{2k}B_{rec}^{2k} \left( \rho_h^{t-k}B_{rec}^{t-k} -1 \right)^2 
    \cdot\ln(2d_h^2)
    \nonumber \\
    &=
    \dfrac{9(t-1)^2 \rho_y^2 \rho_h^2 B_W^2 B_X^2 B_{out}^2 }{ \epsilon^2 \left( \rho_h B_{rec} -1 \right)^2} 
    \bigg( (t-1)(\rho_h B_{rec})^{2t} -2 \rho_h^{t+1}
    \nonumber \\
    & \cdot B_{rec}^{t+1} \dfrac{(\rho_h B_{rec})^{t-1} -1}{\rho_h B_{rec} -1}
    +\rho_h^{2}B_{rec}^{2} \dfrac{(\rho_h B_{rec})^{2t-2} -1}{\rho_h B_{rec} -1}
    \bigg)
    \cdot\ln(2d_h^2)
\end{align}
And so the proof of Proposition~\ref{prop: RNNcover} is complete. $\blacksquare$

\section{Proof of ESN covering number bound}
\label{apx: ESNproof}
Covering an ESN is pretty similar to covering an RNN.
Since $W_{in}$ is random but fixed, we do not need to cover the input layer's output as there is no parameter tuning feature for it.
The same can be said for the recurrent layers.
Therefore, we do not need to define cover sets for those layers as we have done in (\ref{inputLayerCover}), (\ref{firstRecurrentLayerCover}) and (\ref{recurrentLayersCovers}).
The only tunable parameters are those in the output layer, thus we form a $\lambda_t$-cover set $\mathscr{Z}_t$ of $W_{out}H_t$, which its cardinality will be bounded as

\begin{align}
    |\mathscr{Z}_t|=\mathcal{N}\Big\{ W_{out} H_t: W_{out}\in \mathcal{W}_{out},\lambda_t,\|\cdot \|_2 \Big\}\le R
    \label{esnOutputCover}
\end{align}
Where $R$ is equal to the RHS of [Lemma in Bartlett paper].
Now we will fix $W_{out}$ so that we can find a cover element that is close to the output of the ESN.
For every input sequence $\{X_1,X_2,\dots,X_t\}$ we would choose $\hat{Z}_t \in \mathscr{Z}_t$ such that $|Y_t-\sigma_y (\hat{Z}_t)| \le \epsilon_t$. Knowing that 
\begin{align}
    |Y_t-\sigma_y (\hat{Z}_t)| 
    = 
    |\sigma_y (H_t) -\sigma_y (\hat{Z}_t)|
    \le
    \rho_y |H_t - \hat{Z}_t|
    \le
    \rho_y \lambda_t
\end{align}
we only need to set $\lambda_t = \epsilon/\rho_y$ so that we would have a cover element $\hat{Y}_t=\sigma_y (\hat{Z}_t)$ that is in $\epsilon$-proximity of $Y_t$.
We remark that $H_t$ is fixed given a fixed input sequence.
The only thing that determines $R$ in (\ref{esnOutputCover}) is $H_t$'s spectral norm which has already been shown to be upper bounded by the expression in (\ref{HiUpperBound}).
Inserting this into Lemma~\ref{lmma: Bartlett} we would have the metric entropy of an ESN upper bounded by
\begin{align}
    &\ln\left( \mathcal{N}(\mathcal{H}_{ESN,t},\epsilon,\|\cdot\|) \right)
    \le
    \dfrac{\rho_y^2 \|H_t\|^2 B_{out}^2 d_y}{\epsilon^2}\ln (2d_h d_y) \nonumber \\
    &\le
    \dfrac{\rho_y^2 \rho_h^2 B_{in}^2 B_X^2 B_{out}^2 d_y}{\epsilon^2} \left( \dfrac{\rho_h^t B_{rec}^t -1}{\rho_h B_{rec} -1} \right)^2\ln (2d_h d_y)
\end{align}
And this completes the proof of Proposition~\ref{prop: ESNcover}. $\blacksquare$

\end{appendices}


\bibliographystyle{IEEEtran}
\bibliography{IEEEabrv,references}

\begin{thebibliography}{10}
\providecommand{\url}[1]{#1}
\csname url@samestyle\endcsname
\providecommand{\newblock}{\relax}
\providecommand{\bibinfo}[2]{#2}
\providecommand{\BIBentrySTDinterwordspacing}{\spaceskip=0pt\relax}
\providecommand{\BIBentryALTinterwordstretchfactor}{4}
\providecommand{\BIBentryALTinterwordspacing}{\spaceskip=\fontdimen2\font plus
\BIBentryALTinterwordstretchfactor\fontdimen3\font minus
  \fontdimen4\font\relax}
\providecommand{\BIBforeignlanguage}[2]{{%
\expandafter\ifx\csname l@#1\endcsname\relax
\typeout{** WARNING: IEEEtran.bst: No hyphenation pattern has been}%
\typeout{** loaded for the language `#1'. Using the pattern for}%
\typeout{** the default language instead.}%
\else
\language=\csname l@#1\endcsname
\fi
#2}}
\providecommand{\BIBdecl}{\relax}
\BIBdecl

\bibitem{haykin2015cognitive}
S.~{Haykin} and P.~{Setoodeh}, ``Cognitive radio networks: The spectrum supply
  chain paradigm,'' \emph{IEEE Transactions on Cognitive Communications and
  Networking}, vol.~1, no.~1, pp. 3--28, 2015.

\bibitem{chakraborty2016design}
T.~{Chakraborty}, I.~S. {Misra}, and T.~{Manna}, ``Design and implementation of
  {VoIP} based two-tier cognitive radio network for improved spectrum
  utilization,'' \emph{IEEE Systems Journal}, vol.~10, no.~1, pp. 370--381,
  2016.

\bibitem{Rupasinghe2015reinforce}
N.~Rupasinghe and {\.I}.~G{\"u}ven{\c{c}}, ``Reinforcement learning for
  licensed-assisted access of lte in the unlicensed spectrum,'' in \emph{2015
  IEEE Wireless Communications and Networking Conference (WCNC)}.\hskip 1em
  plus 0.5em minus 0.4em\relax IEEE, 2015, pp. 1279--1284.

\bibitem{DSS}
F.~M. {Abinader et al.}, ``Enabling the coexistence of lte and wi-fi in
  unlicensed bands,'' \emph{IEEE Communications Magazine}, vol.~52, no.~11, pp.
  54--61, 2014.

\bibitem{NGA}
``{National 6G Roadmap},'' ATIS NextG Alliance, 02 2022.

\bibitem{RubayetAI}
R.~Shafin, L.~Liu, V.~Chandrasekhar, H.~Chen, J.~Reed, and J.~C. Zhang,
  ``Artificial intelligence-enabled cellular networks: A critical path to
  beyond-5g and 6g,'' \emph{IEEE Wireless Communications}, vol.~27, no.~2, pp.
  212--217, 2020.

\bibitem{chang2018distributive}
H.-H. Chang, H.~Song, Y.~Yi, J.~Zhang, H.~He, and L.~Liu, ``Distributive
  dynamic spectrum access through deep reinforcement learning: A reservoir
  computing-based approach,'' \emph{IEEE Internet of Things Journal}, vol.~6,
  no.~2, pp. 1938--1948, 2018.

\bibitem{hausknecht2015DRQN}
M.~Hausknecht and P.~Stone, ``Deep recurrent {Q}-learning for partially
  observable {MDPs},'' in \emph{2015 aaai fall symposium series}, 2015.

\bibitem{igl2018deep}
M.~Igl, L.~Zintgraf, T.~A. Le, F.~Wood, and S.~Whiteson, ``Deep variational
  reinforcement learning for pomdps,'' in \emph{International Conference on
  Machine Learning}.\hskip 1em plus 0.5em minus 0.4em\relax PMLR, 2018, pp.
  2117--2126.

\bibitem{YueDSA}
Y.~Xu, J.~Yu, and R.~M. Buehrer, ``The application of deep reinforcement
  learning to distributed spectrum access in dynamic heterogeneous environments
  with partial observations,'' \emph{IEEE Transactions on Wireless
  Communications}, vol.~19, no.~7, pp. 4494--4506, 2020.

\bibitem{2013RNNdifficulty}
R.~Pascanu, T.~Mikolov, and Y.~Bengio, ``On the difficulty of training
  recurrent neural networks,'' in \emph{International conference on machine
  learning}.\hskip 1em plus 0.5em minus 0.4em\relax PMLR, 2013, pp. 1310--1318.

\bibitem{huang2019svqn}
S.~Huang, H.~Su, J.~Zhu, and T.~Chen, ``Svqn: Sequential variational soft
  q-learning networks,'' in \emph{International Conference on Learning
  Representations}, 2019.

\bibitem{Hao-Hsuan-DEQN}
H.-H. Chang, L.~Liu, and Y.~Yi, ``Deep echo state {Q}-network ({DEQN}) and its
  application in dynamic spectrum sharing for {5G} and beyond,'' \emph{IEEE
  Transactions on Neural Networks and Learning Systems}, vol.~33, no.~3, pp.
  929--939, 2022.

\bibitem{jaeger2001echo}
H.~Jaeger, ``The “echo state” approach to analysing and training recurrent
  neural networks-with an erratum note,'' \emph{Bonn, Germany: German National
  Research Center for Information Technology GMD Technical Report}, vol. 148,
  no.~34, p.~13, 2001.

\bibitem{tanaka2019RC}
G.~Tanaka, T.~Yamane, J.~B. H{\'e}roux, R.~Nakane, N.~Kanazawa, S.~Takeda,
  H.~Numata, D.~Nakano, and A.~Hirose, ``Recent advances in physical reservoir
  computing: A review,'' \emph{Neural Networks}, vol. 115, pp. 100--123, 2019.

\bibitem{fan2020theoretical}
J.~Fan, Z.~Wang, Y.~Xie, and Z.~Yang, ``A theoretical analysis of deep
  q-learning,'' in \emph{Learning for Dynamics and Control}.\hskip 1em plus
  0.5em minus 0.4em\relax PMLR, 2020, pp. 486--489.

\bibitem{RNNchen}
M.~Chen, X.~Li, and T.~Zhao, ``On generalization bounds of a family of
  recurrent neural networks,'' in \emph{International Conference on Artificial
  Intelligence and Statistics}.\hskip 1em plus 0.5em minus 0.4em\relax PMLR,
  2020, pp. 1233--1243.

\bibitem{POMDPservey}
M.~Lauri, D.~Hsu, and J.~Pajarinen, ``Partially observable markov decision
  processes in robotics: A survey,'' \emph{IEEE Transactions on Robotics},
  2022.

\bibitem{RademacherRL}
Y.~Duan, C.~Jin, and Z.~Li, ``Risk bounds and rademacher complexity in batch
  reinforcement learning,'' in \emph{International Conference on Machine
  Learning}.\hskip 1em plus 0.5em minus 0.4em\relax PMLR, 2021, pp. 2892--2902.

\bibitem{zhang2002}
T.~Zhang, ``Covering number bounds of certain regularized linear function
  classes,'' \emph{Journal of Machine Learning Research}, vol.~2, no. Mar, pp.
  527--550, 2002.

\bibitem{Ganon2020}
L.~Gonon, L.~Grigoryeva, and J.-P. Ortega, ``Approximation bounds for random
  neural networks and reservoir systems,'' \emph{arXiv preprint
  arXiv:2002.05933}, 2020.

\bibitem{boyd2004convex}
S.~Boyd, S.~P. Boyd, and L.~Vandenberghe, \emph{Convex optimization}.\hskip 1em
  plus 0.5em minus 0.4em\relax Cambridge university press, 2004.

\bibitem{foerster2017Replay}
J.~Foerster, N.~Nardelli, G.~Farquhar, T.~Afouras, P.~H. Torr, P.~Kohli, and
  S.~Whiteson, ``Stabilising experience replay for deep multi-agent
  reinforcement learning,'' in \emph{International conference on machine
  learning}.\hskip 1em plus 0.5em minus 0.4em\relax PMLR, 2017, pp. 1146--1155.

\bibitem{reduced-capability}
{3GPP}, ``Technical {S}pecification {G}roup {R}adio {A}ccess {N}etwork; study
  on support of reduced capability {NR} devices ({R}elease 17). {TR} 38.875,''
  3rd Generation Partnership Project (3GPP), Mar. 2021, version 17.0.0.

\bibitem{LTE}
------, ``Evolved universal terrestrial radio access ({E-UTRA}); {P}hysical
  {L}ayer {P}rocedures. {T}echnical {S}pecification ({TS}) 36.213,'' 3rd
  Generation Partnership Project (3GPP), Jun. 2019, version 15.6.0.

\bibitem{wellens2010lessons}
M.~Wellens and P.~M{\"a}h{\"o}nen, ``Lessons learned from an extensive spectrum
  occupancy measurement campaign and a stochastic duty cycle model,''
  \emph{Mobile networks and applications}, vol.~15, no.~3, pp. 461--474, 2010.

\bibitem{hyytia2007rwp}
E.~Hyyti{\"a} and J.~Virtamo, ``Random waypoint mobility model in cellular
  networks,'' \emph{Wireless Networks}, vol.~13, no.~2, pp. 177--188, 2007.

\bibitem{zhu2018improving}
P.~Zhu, X.~Li, P.~Poupart, and G.~Miao, ``On improving deep reinforcement
  learning for pomdps,'' \emph{arXiv preprint arXiv:1804.06309}, 2018.

\bibitem{Bartlett2017}
P.~L. Bartlett, D.~J. Foster, and M.~J. Telgarsky, ``Spectrally-normalized
  margin bounds for neural networks,'' \emph{Advances in neural information
  processing systems}, vol.~30, 2017.

\end{thebibliography}


 





\end{document}